\documentclass[aps,prb,twocolumn,showpacs]{revtex4}
\usepackage[utf8]{inputenc}
\usepackage{bm}
\usepackage{graphicx}
\usepackage{amsmath}
\usepackage{color}
\usepackage{upgreek}
\usepackage{subfigure}
\usepackage{natbib}
\usepackage{booktabs}
\usepackage{fourier} 
\usepackage{array}
\usepackage{makecell}
\usepackage{xcolor}
\usepackage[unicode=true,colorlinks=true,citecolor=blue]{hyperref}
\usepackage{amssymb}

\usepackage{ulem} % need for \sout{ durchgestrichen

\begin{document}

\title{Sign-alternating photoconductivity and magnetoresistance oscillations\\ induced by terahertz radiation in HgTe quantum wells}

\author{M.\,Otteneder,$^1$ I.\,A.\,Dmitriev,$^{1,2}$ S.\,Candussio,$^1$ M.\,L.\,Savchenko,$^{3}$\,D.\,A.\,Kozlov,$^{3}$ V.\,V.\,Bel'kov,$^2$ Z.\,D.\,Kvon,$^{3}$ N.\,N.\,Mikhailov,$^3$ S.\,A.\,Dvoretsky,$^3$ and S.\,D.\,Ganichev$^1$}

\affiliation{$^1$Terahertz Center, University of 
Regensburg, 93040 Regensburg, Germany}

\affiliation{$^2$Ioffe Institute, 
194021 St.\,Petersburg, Russia}

\affiliation{$^3$A.V. Rzhanov Institute of
Semiconductor Physics, Novosibirsk 630090, Russia}

\begin{abstract}

We report on the observation of terahertz radiation induced photoconductivity and of terahertz analog of the microwave-induced resistance oscillations (MIRO) in HgTe-based quantum well (QW) structures of different width. The MIRO-like effect has been detected in QWs of 20\,nm thickness with inverted band structure and a rather low mobility of about $3 \times 10^{5}$~cm$^2$/V\,s. In a number of other structures with QW widths ranging from 5 to 20~nm and lower mobility we observed an unconventional non-oscillatory photoconductivity signal which changes its sign upon magnetic field increase. This effect was observed in structures characterized by both normal and inverted band ordering, as well as in QWs with critical thickness and linear dispersion. In samples having Hall bar and Corbino geometries an increase of the magnetic field resulted in a single and double change of the sign of the photoresponse, respectively. We show that within the bolometric mechanism of the photoresponse these unusual features imply a non-monotonic behavior of the transport scattering rate, which should decrease (increase) with temperature for magnetic fields below (above) the certain value. This behavior is found to be consistent with the results of dark transport measurements of magnetoresistivity at different sample temperatures. Our experiments demonstrate that photoconductivity is a very sensitive probe of the temperature variations of the transport characteristics, even those that are hardly detectable using standard transport measurements.

\end{abstract}

\pacs{}
\maketitle

\section{Introduction}

The richness and diversity of nonequilibrium magnetotransport phenomena in nanostructures make them an indispensable tool for studying their electronic properties, important both for fundamental science and applications. Among recent advances, a particularly intriguing one has been the discovery of the microwave-induced resistance oscillations (MIRO) and associated zero resistance states in two-dimensional electron systems (2DES).\cite{Zudov01,Ye01,Mani02,Zudov03,Yang03,Dorozhkin03,Smet05,Dmitriev2012} Similar to conventional Shubnikov-de Haas (SdH) oscillations in the linear magnetotransport, MIRO are $1/B$-periodic oscillations of the photoresistivity. While the SdH oscillations are governed by the ratio $\mu/\hbar\omega_\mathrm{c}$ of the chemical potential $\mu$ and the cyclotron energy $\hbar \omega_\mathrm{c}=\hbar e B/m_\mathrm{e}$, MIRO reflect the commensurability between the photon energy $\hbar\omega=\hbar 2\pi f$ and the cyclotron energy $\hbar \omega_\mathrm{c}$. Here $B$ is the strength of the magnetic field applied normally to the plane of the 2DES, $f=\omega/2\pi$ is the radiation frequency, and $e$ and $m_\mathrm{e}$ are the electron charge and mass. In the dirty limit $\omega_\mathrm{c}\tau_\mathrm{q}\ll 1$ of strongly overlapping Landau levels, broadened by disorder, and not too strong radiation power $P$, the $B$-dependence of MIRO is well described by the expression\cite{Dmitriev2012}
\begin{equation}\label{MIROeq}
\dfrac{\Delta \rho_{xx}}{\rho_{xx}}= -A \dfrac{\omega}{\omega_\mathrm{c}}\sin\dfrac{2\pi\omega}{\omega_\mathrm{c}}\exp\left(\dfrac{-2\pi}{\omega_\mathrm{c}\tau_\mathrm{q}}\right).
\end{equation}
Here $\Delta \rho_{xx}$ denotes the correction to the dark dissipative resistivity $\rho_{xx}$ induced by the radiation, the amplitude $A$ is proportional to the radiation power $P$, and $\tau_\mathrm{q}$ is the quantum scattering time describing the exponential damping of oscillations at low $B$. The distinctive feature of MIRO is their $\omega/\omega_\mathrm{c}$-periodicity, with the maxima/minima being quarter-of-the-period shifted\cite{Mani04} from the harmonics of the cyclotron resonance (CR) at integer $\omega/\omega_\mathrm{c}$.

The exponential damping of MIRO at low $B$, see Eq.~(\ref{MIROeq}), imposes a rigid 
necessary condition for their observation, 
\begin{equation}\label{ftq}
f\tau_\mathrm{q}\gtrsim 1\,,
\end{equation}
which implies that the illumination frequency $f$ should be at least of the same order of magnitude as the quantum scattering rate $1/\tau_\mathrm{q}$. Since this scattering rate in most materials falls into the picosecond or sub-picosecond range, it is not surprising that strong microwave induced resistance oscillations and, especially, zero resistance states have so far been observed exclusively in 2DES of highest quality, namely, in ultra-high mobility GaAs/AlGaAs heterostructures,\cite{Zudov01,Ye01,Mani02,Zudov03,Yang03,Dorozhkin03,Smet05,Dmitriev2012} in ultra-clean 2DES on the surface of liquid helium,\cite{Konstantinov2009,Konstantinov2010,Yamashiro2015,Zadorozhko2018} and, more recently, also in the highest quality Si/SiGe\cite{Zudov2014,Shi2014} and MgZnO/ZnO\cite{Karcher2016} heterostructures.

\begin{table*}[tb]
	\centering
\begin{tabular*}{\textwidth}{c@{\extracolsep{\fill}}c c c c c c c}
	\hline
	\hline
	\makecell{sample\\ number} & \makecell{QW thickness\\ $d$\,(nm)} & geometry & \makecell{substrate\\ orientation} & $n_\mathrm{s}\,\left(10^{11}\,\text{cm}^{-2}\right)$ & $\mu\,\left(10^{4}\,\text{cm}^2/\text{Vs}\right)$ & size (mm)\\ 
	\hline
	\hline 
	\rule{0pt}{3ex} 
	\#1 %100623
	& 20 & Hall bar with gate & (001)& $1\div 7$ & $17\div 50~$ & $l=0.25$, $w=0.05$\\
	\#2 %100623
	& 20 & van der Pauw & (001)& 1 & 17 & $a=5$ \\
	\#3 %130213
	& 20 & Hall bar & (013) & $8.2$ & $15$ & $l=0.25$, $w=0.05$ \\
	\#4 %061222
	& 8 & Hall bar & (013) & $7.5$ & $6.7$ & $l=0.25$, $w=0.05$\\
	\#5 %130211
	& 6.6 & Corbino & (013) & $12$  & $4.4$ & $r_\mathrm{i}=0.36$, $r_\mathrm{o}=1.9$\\
	\#6 %151209
	& 6.5 & Hall bar & (013) & $10$ & $13$ & $l=0.25$, $w=0.05$\\
	\#7 %151209
	& 6.5 & Corbino & (013) & $9.7$ & $11$ & $r_\mathrm{i}=0.25$, $r_\mathrm{o}=1.9$\\
	\#8 %130411
	& 5.0 & Corbino & (013) & $4.0$ & $0.14$ & $r_\mathrm{i}=0.32$, $r_\mathrm{o}=1.2$ \\
	\hline
	\hline
\end{tabular*}
\caption{Parameters of the samples used for presentation in Figs.~\ref{MIRO}-\ref{HallTr}. Electron densities and mobilities were obtained from low temperature magnetotransport measurements. Note that the actual mobilities for Corbino samples can have slightly larger values than those given in the table due to unavoidable contribution of the contact resistance in the corresponding two-point measurements.}
\label{sampledata}
\end{table*}
 
On the other hand, in the terahertz range of frequencies the requirement (\ref{ftq}) on the material quality becomes less restrictive, which opens a possibility to study MIRO and related effects in a wider range of materials and conditions. 
Indeed, recent experiments\cite{Hermann2016,Hermann2017} have shown that in the terahertz range pronounced MIRO-like oscillations can be induced in GaAs/AlGaAs structures with mobility as low as $1.5 \times 10^{5}$~cm$^2$/V\,s, which is two orders of magnitude smaller than in conventional MIRO studies. In particular, using THz frequencies allows one a search for MIRO in topological materials with strong spin-orbit coupling and inverse band ordering, but so far rather low mobilities. Such studies could provide additional information on specific transport mechanisms in these materials, and is also important for a deeper understanding of the MIRO effect itself.

Here we report on the observation of MIRO-like oscillations excited by radiation with a frequency of $f = 0.69$~THz in a HgTe/CdHgTe heterostructure hosting a quantum well (QW) with inverted band structure. MIRO-like $1/B$-oscillations coupled to CR and its harmonics have been detected in a single sample with QW of 20~nm thickness. The corresponding results are presented in Sec.~\ref{secMIRO} after introducing our samples characteristics and the experimental setup in Sec.~\ref{secSAMPLES}.

A particular feature of the HgTe/CdHgTe heterostructures is that one can obtain  a Dirac-like, inverted and normal energy dispersions without changing the material.\cite{Bernevig2006}
%~\cite{Bernevig2006,Koenig2007,Dyakonov1981,Kvon2008,Diehl2009,Wittman2010,Ikonnikov2010,Buettner2011,Kvon2011,Hancock2011,Ikonnikov2011,Zholudev2012,Ikonnikov2012,Kozlov2012,Olbrich2013,Shuvaev2013,Shuvaev2013a,Orlita2014,Zoth2014,Dantscher2015,Dantscher2017} 
Thus, investigating various electronic properties in HgTe-based quantum wells with different thicknesses one can address similarities and differences of emerging phenomena for different types of electron energy spectra. While searching for MIRO-like oscillations in heterostructures corresponding to all three types of energy dispersion, we unexpectedly found a non-oscillatory photoconductivity signal which, however, changes its sign 
upon variation of the magnetic field. The effect is observed in structures with different QW thicknesses characterized by normal and inverted band ordering, as well as in QWs with critical thickness corresponding to the linear energy dispersion. We show that in samples having Hall bar and Corbino geometries an increase of the magnetic field results in a single and double change of the sign of the photoresponse, respectively. The details of this unusual sign-alternating 
photoresponse are presented in Sec.~\ref{secINVERSION}. In Sec.~\ref{secDISCUSSION} the origin of the sign inversions is discussed within the bolometric mechanism of the photoresponse. This model explains well the experimental findings, and is additionally supported by the temperature dependence of the dark magnetotransport.
Main results are summarized in Sec.~\ref{secSUMMARY}.  

\section{Samples and experimental setup}
\label{secSAMPLES}

For our experiments we used HgTe/CdHgTe QWs grown by molecular beam epitaxy on GaAs substrates with (013) and (001) orientation.\cite{Dvoretsky2010}
We studied a number of samples with different band order and dispersion, including (i)
QWs with thickness $d=20$ and 8\,nm (inverted band order and parabolic dispersion),\cite{Kvon2008,Zoth2014} 
(ii) QWs with $d=5.7$\,nm (normal band order and parabolic dispersion), and 
(iii) QWs with $d=6.6$ and 6.5\,nm close to critical thickness,~\cite{Bernevig2006,Buettner2011,Kvon2011} 
with the band structure described by a linear dispersion. Geometrical and transport parameters of all samples are summarized in Table\,\ref{sampledata}.

For photoconductivity, photoresistivity, and magnetotransport measurements we used samples having both Corbino and Hall bar geometries. Hall bars having lengths $l$, widths $w$, and with potentiometric contacts spaced by 250~$\upmu$m were fabricated using optical lithography. Sample \#1 has a semitransparent gate composed of a 20~nm Ti layer and a 5~nm Au layer, which was made on top of a 200~nm-thick SiO$_2$ layer deposited using the plasmochemical method. Both center and outer ring contacts in Corbino samples were made on top of the HgTe-based heterostructure via diffusion of indium. The inner, $r_\mathrm{i}$, and outer radii, $r_\mathrm{o}$, of the resulting Corbino samples are given in Table\,\ref{sampledata}. Furthermore, for determination of the CR positions we performed measurements of radiation transmission. For these measurements we used large area samples having either Corbino or van der Pauw geometry. The van der Pauw samples are square-shaped with a side length $a$.

The samples were placed in an optical temperature-regulated cryostat with $z$-cut crystal quartz windows. All windows were covered by a thick black polyethylene film which is transparent in the terahertz range but prevents uncontrolled illumination of the sample with room light in both visible and infrared ranges. %To prevent uncontrolled illumination of the sample with room visible and infrared light, all windows were covered by a thick black polyethylene film which is transparent in the terahertz range but blocks both visible and infrared light. 
We used a continuous flow cryostat, which allows us to measure the temperature dependence of the photoresponse. The magnetic field ${\bm B}$ up to 7~T has been applied  normal to  the QW plane. 
We also performed standard  4-terminal magnetotransport measurements in the absence of irradiation in order to confirm the origin of the sign-alternating photoresponse and to determine the electron densities and mobilities for all investigated samples (see Table \ref{sampledata}).
%For determination of electron densities and mobilities as well as for analysis of the origin of the  sign inversions in the photoresponse we additionally measured magnetotransport in the absence of irradiation.
% using standard 4-terminal setup.
% For that purpose we used a 4-terminal configuration and applied low-frequency ($f_\mathrm{ac}=12\,$Hz) alternating currents of $I \approx 10^{-7}\,$A.  
%Table \ref{sampledata} shows the carrier densities and mobilities for all investigated samples.

Terahertz radiation in our experiments is provided by a continuous wave terahertz molecular gas laser\cite{DMSPRL09,Olbrich_PRB_11} optically pumped by a CO$_2$ laser.
We used CH$_3$OH and CH$_2$O$_2$ as active media to obtain linearly polarized radiation with a frequency of $f=2.54\,$, 1.63, and $0.69\,$THz corresponding to photon energies $\hbar\omega=10.50$, 6.74, and 2.85\,meV, respectively.
The peak power $P_{\rm R}$ at a sample position was in the range from 5 to 30\,mW. The beam shape and parameters were
controlled using a pyroelectric camera.\cite{schneider2004} The laser spot diameter at FWHM was 0.15, 0.3, and 0.45~mm
for laser lines with $f=2.54\,$ 1.63, and $0.69\,$THz, respectively. These values have been used to obtain peak intensities $I_{\rm R}$.
%The corresponding wavelengths are 118, 184 and 432\,$\mu$m, respectively. 
%The beam shape and parameters were controlled using a pyroelectric camera.\cite{schneider2004}
%For $f=2.54\,$THz and $f=1.63\,$THz, the beam incident on the sample had spot size of about 2\,mm and power $P\approx 25$\,mW. For $f=0.69\,$THz, the spot size was about 3\,mm and power was $\approx 7$\,mW. 
In some measurements, quarter wave plates made of $x-$cut crystal quartz have been used to obtain right-handed ($\sigma^+$) and left-handed ($\sigma^-$) circularly polarized radiation. 

\begin{figure}[t]
\begin{center}
\includegraphics[width=\linewidth]{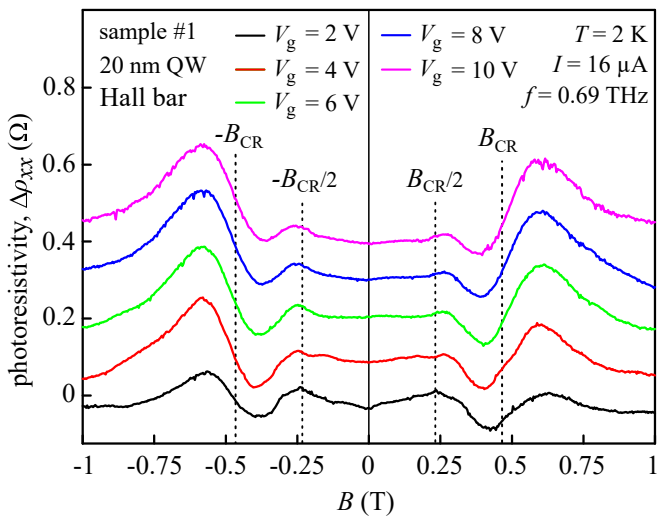}
\caption{Magnetic field dependence of the photoresistivity measured in a Hall bar sample with $d =20$~nm (sample \#1) at $T=2$~K under illumination with linearly polarized radiation at $f=0.69~$THz. The data were obtained using the double modulation technique with an applied alternating current of $I = 16~\upmu$A. The dashed vertical lines indicate the position of CR, $B_\text{CR}=0.47~$T, and of its second harmonics, $B_\text{CR}/2$. The value of $B_\text{CR}$ was obtained from the transmission measurements (not shown) in a square-shaped van der Pauw sample made from the same wafer (sample \#2). The presented data correspond to different gate voltages between $V_\mathrm{g}=2~$V and $V_\mathrm{g}=10~$V. The intensity of the terahertz radiation incident on the sample was $I_\mathrm{R}=0.06~$W/cm$^2$.
%... $P_\text{powermeter}=7~$mW.
Traces for different gate voltages are marked by different colors and are vertically shifted by 0.1$~\Omega$ for visibility. 
}
\label{MIRO}
\vspace{-1cc}
\end{center}
\end{figure}

\begin{figure}[t]
\begin{center}
\includegraphics[width=\linewidth]{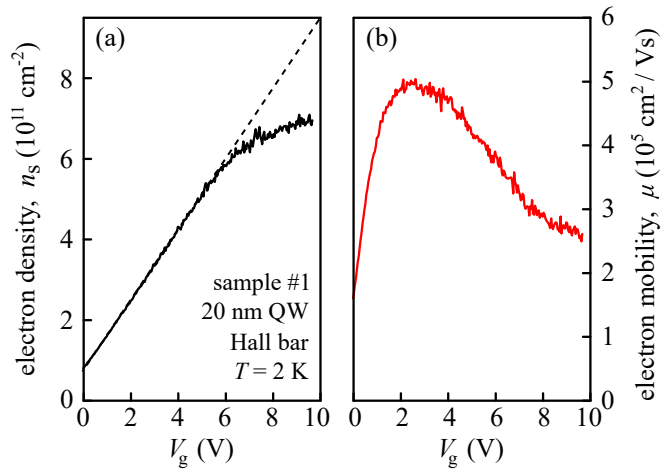}
\caption{The electron density $n_\mathrm{s}$ [panel (a)] and mobility $\mu$ [panel (b)] as a function of the applied gate voltage $V_\mathrm{g}$ for Hall bar sample \#1. Both dependencies are extracted from the magnetotransport measurements  without terahertz irradiation, as described in the text. At high $V_\mathrm{g}>6$~V the dependence $n_\mathrm{s}(V_\mathrm{g})$ starts to deviate from linear (dashed line), which is most probably related to occupation of the second electron subband in the QW.\cite{dobretsova2018}}
\label{MIROtr}
\vspace{-1cc}
\end{center}
\end{figure}

The laser beam was chopped at a frequency $f_\text{chop}=170\,$Hz and focused inside the optical cryostat by a parabolic mirror. Photoconductivity was measured using either the dc-photoconductivity or double-modulation setup. In the former setup, a dc bias voltage or current is applied to the sample and the photosignal modulated at the chopper frequency $f_\text{chop}$ is measured as the voltage drop over a load resistor using standard lock-in technique. The photoconductivity and photocurrent signals can be extracted by subtracting (summing up) the signals obtained for opposite dc-bias polarities. In the double modulation setup, an alternating current is applied to the sample. The photosignal measured over two contacts is modulated at the ac frequency $f_\mathrm{ac}=12$\,Hz as well as at the chopper frequency $f_\text{chop}$ of the incident laser beam, and is read out by two lock-in amplifiers in series.

\section{MIRO-like oscillations}
\label{secMIRO}

MIRO-like oscillations were observed in the sample having the highest mobility (sample \#1, see Tab.~\ref{sampledata}). This sample hosts a 20~nm HgTe QW and is equipped with a semitransparent gate. The sample was illuminated by linearly polarized radiation with a frequency of $f=0.69~$THz. Figure \ref{MIRO} presents the magnetic field dependence of the terahertz radiation induced change in resistivity obtained at a temperature of $T=2$~K.
% and an applied current of $I=16~\mu$A {\color{red}which results in a heating of the sample and, therefore, a suppression of Shubnikov-de Haas oscillations. Maxim} 
The data clearly demonstrate that terahertz radiation results in an oscillatory photoresistivity. The nodes of the oscillatory signal coincide with CR at $B=B_\mathrm{CR}=2\pi m_\mathrm{e} f/e$ and with its second harmonics at $B=B_\mathrm{CR}/2$, as indicated by the vertical dashed lines. This observation 
%{\color{red}together with the fact that the oscillatory behavior is independent from the value of the gate voltage (Dima Kozlov, I disagree.)} 
provides a clear evidence that the oscillations are indeed of the same origin as MIRO observed before, see Eq.~(\ref{MIROeq}). Similar behavior has also been obtained at $T=4.2$~K. Note that the CR position, $B_\text{CR}=0.47$~T (corresponding to the cyclotron mass $m_\mathrm{e}=0.019 m_0$), was obtained from independent measurements of the terahertz radiation transmission (not shown). The transmission measurements were performed using circularly-polarized terahertz radiation and displayed conventional CR dips at positive (negative) $B$ for right-handed (left-handed) polarization.  

\begin{figure}[t]
\begin{center}
\includegraphics[width=\linewidth]{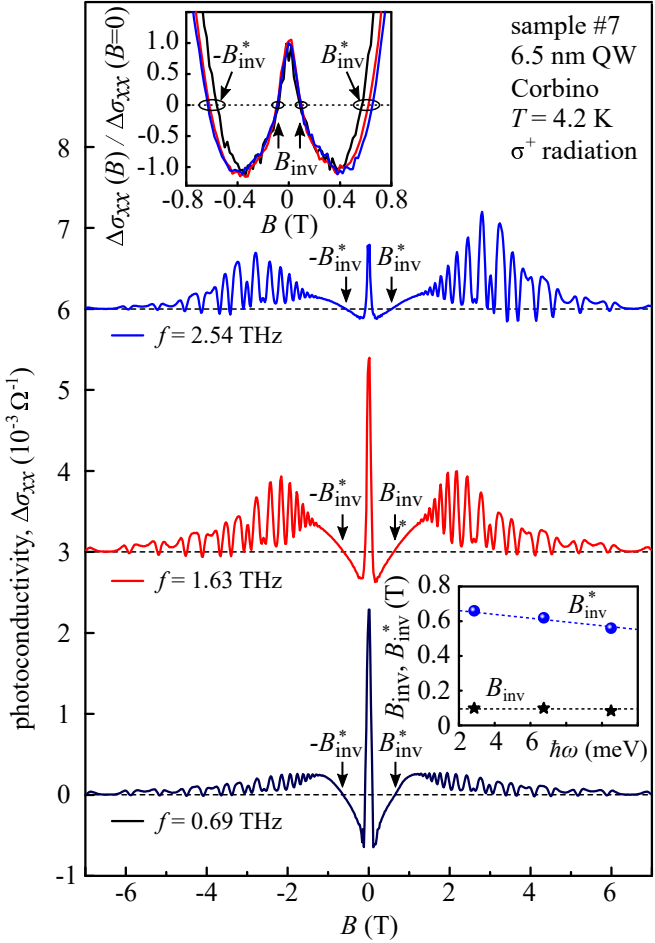}
\caption{Magnetic field dependence of the  photoconductivity $\Delta \sigma_{xx}$ measured in Corbino sample~\#7 ($d=6.5$~nm) illuminated by right-handed circularly polarized radiation of frequencies $f=2.54,$ 1.63, and 0.69~THz.
%with the intensity $I_\mathrm{R}=xx~$W/cm$^2$... $P_\text{power meter}=10$, 26 and 4~mW, respectively. 
The two upper curves are vertically shifted each by 3$\times 10^{-3}\Omega^{-1}$ for visibility. The photoconductivity exhibits two sign inversions, going from positive to negative values at $B=B_{\rm inv}$ and back to positive values at $B=B_{\rm inv}^*$. The top inset shows the low-field portion of the $B$-dependence of the photoconductivity for all three frequencies (here the data are normalized to the values of photoconductivity at $B=0$). The bottom inset shows the dependencies of the first ($B_{\rm inv}$) and second ($B_{\rm inv}^*$) inversion points on the photon energy $\hbar \omega$.
}
\label{PcondCorb}
\vspace{-1cc}
\end{center}
\end{figure}

The oscillations in Fig.~\ref{MIRO} are detected in a wide electron density range controlled by the applied gate voltage varying from $V_\mathrm{g}=2$ to 10 V. The corresponding changes in the electron density and mobility, as obtained from magnetotransport measurements, are shown in Fig.~\ref{MIROtr}. 
The density $n_\mathrm{s}(V_\mathrm{g})$  in Fig.~\ref{MIROtr}(a) was obtained from the Hall resistance $\rho_{xy}$ measured as a function of the applied gate voltage $V_\mathrm{g}$ at fixed $B=0.2$~T. The linearity of $\rho_{xy}$ as a function of $B$ was checked in the whole relevant interval of gate voltages, $0<V_{g}<10$~V. Using the obtained $n_\mathrm{s}(V_\mathrm{g})$, the electron mobility $\mu(V_\mathrm{g})$ [panel (b)] was extracted from the measured diagonal resistivity at $B=0$.
%as $\mu(V_\mathrm{g})=[e n_\mathrm{s}(V_\mathrm{g}) \rho_{xx}(V_\mathrm{g})]$.
It is seen that in the range $0<V_{g}<10$~V the electron density increases from 2 to 7$\times 10^{11}$~cm$^{-2}$, while the mobility decreases from 5 to 2.5$\times 10^{5}$~cm$^{2}$/Vs.
Figure \ref{MIRO} reveals that these changes result in a moderate increase of the MIRO amplitude. Note that the positions of the MIRO nodes (and oscillation minima/maxima as well) do not vary with the electron density, in agreement with Eq.~(\ref{MIROeq}).

\begin{figure}[t]
\begin{center}
\includegraphics[width=\linewidth]{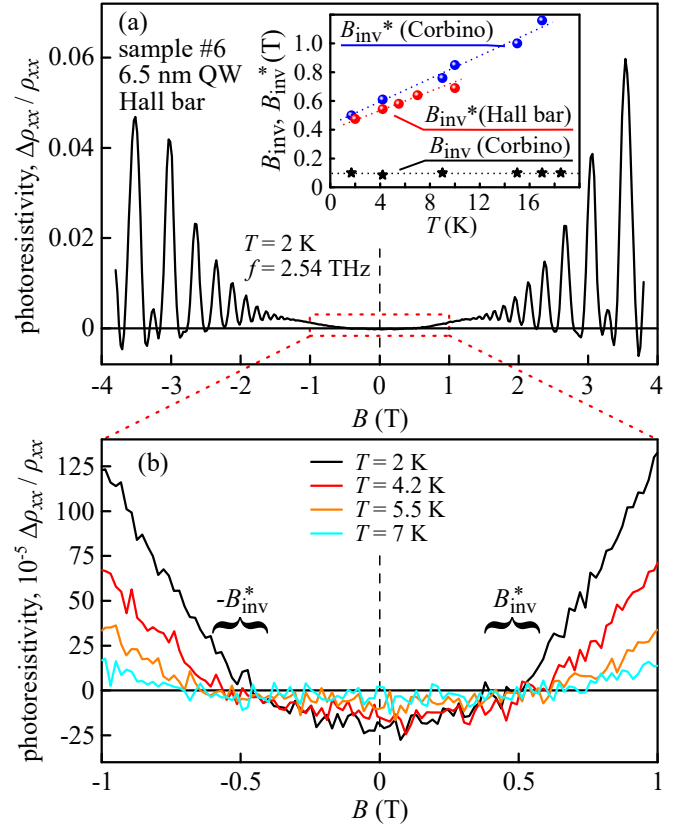}
\caption{
Magnetic field dependence of the photoresistivity (normalized to the dark resistivity $\rho_{xx}$) measured in the Hall bar sample~\#6 ($d=6.5$~nm) under incidence of linearly polarized radiation of frequency $f = 2.54$~THz and intensity $I_\mathrm{R}=1.2~$W/cm$^2$.
%.. $P_\text{powermeter}=35$~mW}. 
Panel (a) shows the data for $T=2~$K in the full range of magnetic fields, while panel (b) shows the  range of $B$ corresponding to the sign inversions and presents data obtained at several temperatures. The photoresistivity goes from negative to positive values at $|B|=B_{\rm inv}^*$. The inset in panel (a) shows the $T$-dependence of $B_{\rm inv}^*$. Note that the values $B_{\rm inv}^*$ nearly coincide with positions of the second inversion points in the Corbino sample~\#7 made from the same wafer, see Fig.~\ref{PcondCorb}.}
\label{PcondHall}
\vspace{-1cc}
\end{center}
\end{figure}

While MIRO have been typically detected in high-mobility samples with $\mu\gtrsim 10^6$~ cm$^2/$Vs, see Ref.~\onlinecite{Dmitriev2012}, the mobility of our sample~\#1 is substantially lower. A short momentum relaxation time $\tau$ (in the range from 2.8 to 5.7 ps for the mobility range quoted above) implies an even shorter quantum lifetime $\tau_\mathrm{q}<\tau$. This results in a relatively small value of $f\tau_\mathrm{q}\sim 1$, see Eqs.~(\ref{MIROeq}) and (\ref{ftq}), and causes a strong decay of the observed oscillations. At the same time, previous theoretical and experimental studies (see e.g. Refs.~\onlinecite{Dmitriev2012,Hermann2017}) indicate that the amplitude of MIRO rapidly decreases with the raise of radiation frequency, consistent with the theoretical $\omega^{-4}$ scaling of $A$ in Eq.~(\ref{MIROeq}). It is thus not surprising that at the higher frequency $f=2.54$~THz MIRO were not detected for available intensities of continuous illumination, despite an obviously larger value of $f\tau_\mathrm{q}$. At this higher frequency we detected solely a resonant photoresponse in the vicinity of $B_\mathrm{CR}$ (not shown), which is attributed to the bolometric CR signal (the change of the conductivity due to resonant electron gas heating in the vicinity of CR). 
Note that, by contrast, at $f=0.69$\,THz no discernible CR-induced bolometric response is seen in Fig.\,\ref{MIRO}.

\section{Sign-alternating terahertz photoconductivity and photoresistivity}
\label{secINVERSION}

\begin{figure}[t]
\begin{center}
\includegraphics[width=\linewidth]{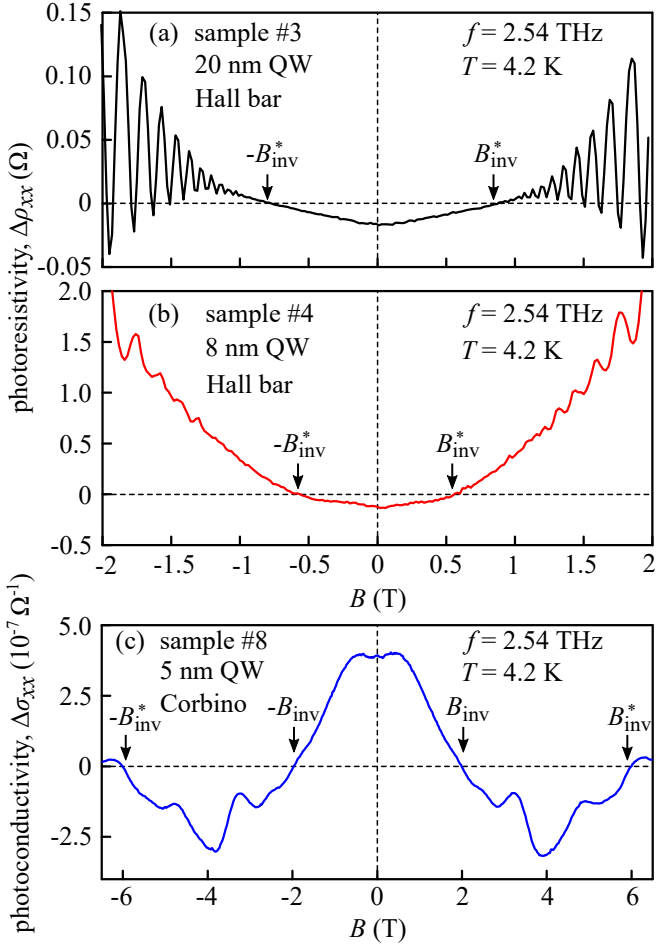}
\caption{Photoresponse for different samples as a function of magnetic field. The data were obtained at $T=4.2$~K under irradiation with $f=2.54~$THz and $I_\mathrm{R}\simeq 0.7~$W/cm$^2$.
%$P_\text{powermeter}=20$~mW}.  
Panels (a) and (b) show photoresistivity $\Delta\rho_{xx}$ for samples \#3 ($d=20$~nm) and \#4 ($d=8$~nm) having Hall bar geometry. Panel (c) displays the  photoconductivity $\Delta\sigma_{xx}$ for the Corbino sample \#8 with $d=5~$nm. Note that the sample \#8 has much smaller mobility in comparison to other samples, see Table~\ref{sampledata}. Therefore, both $B_{\rm inv}$ and  $B_{\rm inv}^*$ are shifted to stronger fields.}
\label{Pcondother}
\vspace{-1cc}
\end{center}
\end{figure}

\begin{figure}[t]
\begin{center}
\includegraphics[width=\linewidth]{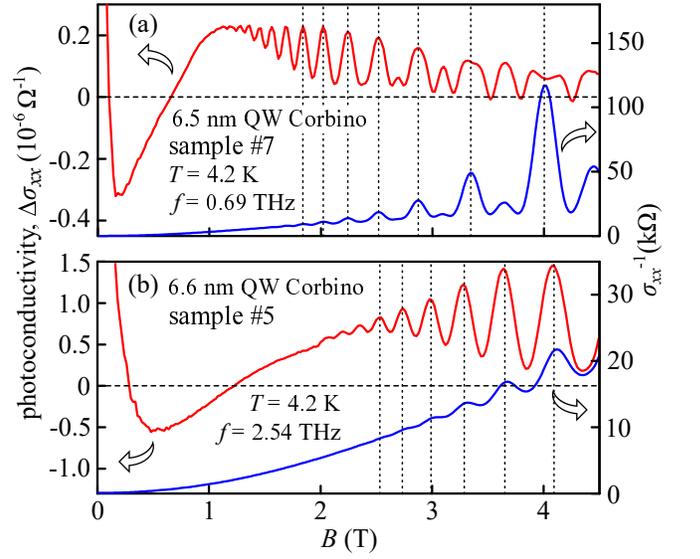}
\caption{Magnetic field dependencies of the photoconductivity $\Delta \sigma_{xx}$ induced by linearly polarized THz radiation and of the reciprocal conductivity $\sigma_{xx}^{-1}$ obtained in magnetotransport measurements without THz irradiation. Panel (a) shows data for Corbino sample \#7 illuminated by radiation with $f=0.69$~THz and intensity $I_\mathrm{R}=0.05~$W/cm$^2$,
%$P_\text{powermeter}=6$~mW)}
while panel (b) presents data for Corbino sample~\#5 illuminated by radiation with $f=2.54$~THz and intensity $I_\mathrm{R}=1~$W/cm$^2$.
%$P_\text{powermeter}=30$~mW)}.
}
\vspace{-1cc}
\label{PcondVSTr}
\end{center}
\end{figure}

In search for MIRO in HgTe-based QWs we performed measurements on many samples with different QW thickness, band ordering, and geometry, see Table\,\ref{sampledata}. So far, MIRO were detected exclusively in sample~\#1 with the highest mobility as presented above. In all other samples \#3-\#8 we instead observe an unusual sign-alternating photoresponse at low $B$, which is followed at higher $B$ by conventional $1/B$-periodic oscillations with the period corresponding to that of SdH oscillations, see Figs.\,\ref{PcondCorb}%=,PcondHall,PcondVSTr,PcondBcr,
-\ref{Pcondother}. A characteristic example of such behavior is shown in Fig.\,\ref{PcondCorb}. 
It displays photoconductivity (the terahertz radiation-induced change $\Delta\sigma_{xx}$ of the longitudinal conductivity $\sigma_{xx}$) measured in sample~\#7 having Corbino geometry. This sample hosts a QW of width $d=6.5$\,nm close to the critical thickness and, consequently, is characterized by an almost linear energy dispersion.\cite{Bernevig2006,Buettner2011,Kvon2011}

The high-field portion of the data in Fig.\,\ref{PcondCorb} exhibits pronounced magneto-oscillations related to Shubnikov-de Haas oscillations, which will be discussed in detail below. Strikingly, in addition to these expected oscillations we observe a double sign inversion of the photoconductivity in the low magnetic field region, substantially before the quantum oscillations set in. This double sign inversion was detected for different radiation frequencies, see Fig.~\ref{PcondCorb}. The positions of the first ($B_{\rm inv}$) and second ($B_{\rm inv}^*$) inversion points behave differently upon variation of the photon energy (see top and bottom insets of Fig.\,\ref{PcondCorb}). While the first inversion is observed to be independent of frequency, the second one slightly shifts to lower fields for larger photon energies $\hbar\omega$. We found that the inversion points always lie symmetrically with respect to $B=0$. In the following, $B_{\rm inv}$ and $B_{\rm inv}^*$ refer to their absolute values. 

In contrast to Corbino samples, in Hall bar geometry the low-$B$ photoresponse features a single sign inversion only, as exemplified in Fig.\,\ref{PcondHall} for a sample~\#6 made from the same wafer as sample~\#7. In Hall bars, one measures photoresistivity (the terahertz radiation-induced change $\Delta\rho_{xx}$ of the longitudinal resistivity $\rho_{xx}$) rather than photoconductivity. The figure shows that photoresisitivity is negative in the low-$B$ region and changes sign at $B=B_{\rm inv}^*$, which corresponds to the positions of the second inversion points in the Corbino sample~\#7 measured under the same conditions. Figure \ref{PcondHall}(b) and the inset of Fig.\,\ref{PcondHall}(a) present the temperature behavior of the inversion point $B_{\rm inv}^*$ demonstrating that its position moves to higher $B$ with rising temperature. The same temperature dependence of the second inversion point $B_{\rm inv}^*$ is observed in Corbino geometry sample~\#7 as shown in the inset to Fig.\,\ref{PcondHall}. Taken together, these observations demonstrate that the origin of this sign inversion is the same in both geometries. By contrast, the first inversion point $B_{\rm inv}$ is observed in Corbino geometry only, and does not depend both on either temperature or radiation frequency. 

\begin{figure}[t]
\begin{center}
\includegraphics[width=\linewidth]{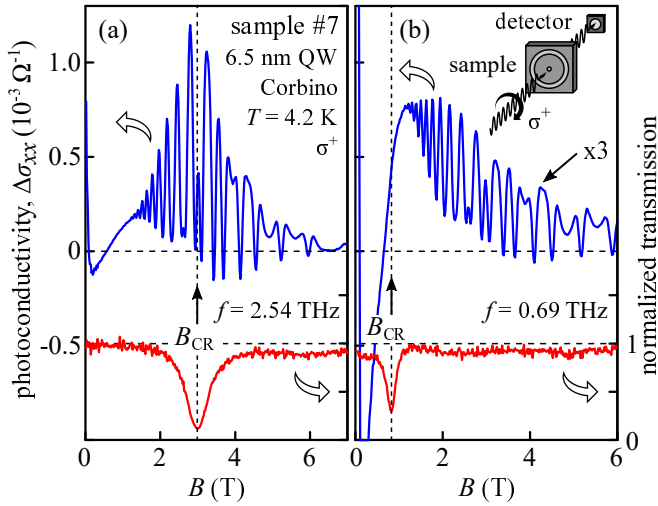}
\caption{Magnetic field dependencies of the photoconductivity $\Delta \sigma_{xx}$ and of the radiation transmission obtained for Corbino sample~\#7 illuminated with right-handed circularly polarized radiation of frequencies $f=2.54$~THz [panel (a)] and 0.69~THz [panel (b)] and intensities $I_\mathrm{R}=0.3$ and $0.06$ W/cm$^2$,
% $P_\text{powermeter}=10$~mW and 7~mW
respectively. Vertical dashed lines indicate positions $B_\text{CR}$ of CR. Note that the photoconductivity values in panel (b) are multiplied by a factor of 3.
% for better comparability.
}
\label{PcondBcr}
\vspace{-1cc}
\end{center}
\end{figure}

Similar results were obtained in other samples having different energy dispersion, namely, in samples with inverted band order [$d=20$ and 8 nm, see Fig.~\ref{Pcondother} (a) and (b)], linear dispersion [$d=6.5$~nm, see Fig.~\ref{PcondVSTr} (b)], and normal band order [$d=5$~nm, Fig.~\ref{Pcondother} (c)]. The only difference between data for all these samples was in the positions of $B_{\rm inv}$ and $B_{\rm inv}^*$. The origin of the observed sign-alternating photoresponse at low $B$ is discussed below in Sec.~\ref{secDISCUSSION}.

Now we turn to the oscillatory part of the photoconductivity at higher magnetic fields, see Figs.\,\ref{PcondCorb} - \ref{Pcondother}.
%, \ref{PcondHall}, \ref{PcondVSTr}, \ref{PcondBcr}, and 
In Fig.\,\ref{PcondVSTr} we show the photoconductivity signals measured on two Corbino samples with QW widths close to the critical thickness, namely, on sample\,\#7 ($d = 6.5$\,nm) and sample\,\#5 ($d = 6.6$\,nm). Here we compare the photoconductivity with the magnetotransport data (the inverse $1/\sigma_{xx}$ of the longitudinal conductivity) measured in the absence of terahertz radiation. The data reveal that the positions of minima and maxima in $\Delta \sigma_{xx}$ coincide with those of the Shubnikov-de Haas oscillations in $1/\sigma_{xx}$. The coincidence of the photoresponse oscillations and SdH has been detected for all samples. Such behavior is a typical manifestation of the radiation-induced heating of electrons leading to suppression of the SdH oscillations.\cite{Zoth2014}  
At higher magnetic fields, the form of the SdH oscillations in Fig.\,\ref{PcondVSTr} is more complicated since both cyclotron and spin gaps become resolved. 

Studies of quantum oscillations at different frequencies have shown that for high photon energies the oscillating signal is substantially enhanced at the positions of CR. An example of such resonant photoresponse is shown in Fig.~\ref{PcondBcr}(a). Note that the position of CR was determined from simultaneous transmission measurements. Surprisingly, no such enhancement is observed at lower frequency ($f=0.69$\,THz), where the position of CR lies below the magnetic field region of magneto-oscillations. Indeed, as demonstrated in Fig. ~\ref{PcondBcr}(b), in this case the photoresponse in the region of CR is apparently completely featureless.

\section{Discussion}
\label{secDISCUSSION}

\begin{figure}[t]
\begin{center}
\includegraphics[width=\linewidth]{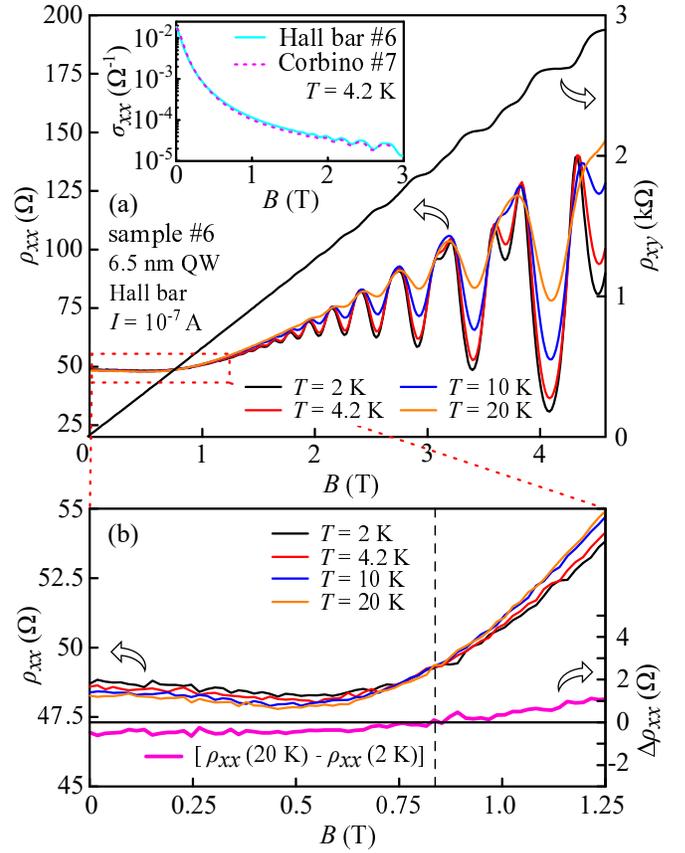}
\caption{Magnetotransport data obtained on Hall bar sample~\#6 with $d= 6.5$~nm. Panel (a) displays the sheet resistance $\rho_{xx}$ for four different temperatures in the range from 2~K up to 20~K [see legend in panel (b)], as well as the Hall resistance $\rho_{xy}$ measured at $T=2$~K. Panel (b) presents a zoom of the magnetotransport data at small magnetic fields. The lowest curve displays the temperature variation of $\rho_{xx}$ between $T=20$~K and 2 K (right axis). In the inset, a comparison of the conductivity $\sigma_{xx}$ is displayed for Corbino and Hall bar samples made from the same wafer. The longitudinal conductivity of the Hall bar is obtained via tensor inversion using the measured longitudinal and Hall resistivities.}
\label{HallTr}
\vspace{-1cc}
\end{center}
\end{figure}

In this section we discuss our main unexpected experimental result: we systematically observe a double change of the sign of photoconductivity in Corbino geometry and a single inversion of photoresistivity in Hall bar geometry in a number of investigated samples having different energy dispersion.

We begin with the discussion of the first (low-field) inversion point $B_{\rm inv}$ in Corbino geometry, which was shown to be independent of both temperature and photon energy, see insets in Figs.~\ref{PcondCorb} and \ref{PcondHall}. Within the semiclassical Drude description of the bolometric effect, such inversion is expected to necessarily occur as a result of the transition from classically weak to classically strong magnetic fields at $\omega_\mathrm{c}\tau=1$. Here $\tau$ is the transport scattering time. Indeed, within the Drude model the dc conductivity $\sigma_{xx}\propto \tau/(1+\omega_\mathrm{c}^2\tau^2)$ is proportional to $\tau$ in the parametric region $\omega_\mathrm{c}=e B/m_\mathrm{e}\ll 1/\tau$, while in the opposite limit $\omega_\mathrm{c}\gg 1/\tau$ the conductivity $\sigma_{xx}\propto 1/\tau$. Therefore, assuming that the transport time $\tau$ is modified under illumination (i.e., due to heating of carriers by the terahertz electric field), the sign of the photoconductivity should be opposite in these two regions. Within such a conventional bolometric mechanism the inversion field $B_{\rm inv}$ is thus given by
\begin{equation}\label{Binv}
B_{\rm inv}=\frac{m_\mathrm{e}}{e\tau}\ .
\end{equation}
To check whether the position of the first inversion point $B_{\rm inv}$ indeed follows Eq.~(\ref{Binv}), we extracted the transport lifetime from the dark transport measurements and the cyclotron mass $m_\mathrm{e}$ from the position of CR in the transmission data.
For higher accuracy we used times $\tau$ extracted from the transport measurements carried out on Hall bar samples made from the same wafer as the Corbino structures.
The obtained values of $m_\mathrm{e}$ and $\tau$ accurately reproduce the position of the first inversion point, which supports the bolometric nature of the observed photoresponse at low magnetic fields. Also in agreement with our observations, within this mechanism the position of the inversion point $B_{\rm inv}$ should be independent of the radiation frequency and temperature (as long as the variation of $\tau$ due to heating remains small) since the change of frequency can only modify the magnitude of heating.

While the inversion of the photoconductivity sign at $B_{\rm inv}=m_\mathrm{e}/e\tau$ is not surprising, the absolute sign of the photoinduced signal on both sides of $B_{\rm inv}$ is opposite to what one would ordinarily expect. Indeed, the electron gas heating usually reduces the transport time due to an additional contribution of the electron-phonon scattering to the momentum scattering rate.

In this case, the conductivity should be suppressed by illumination for $|B|<B_{\rm inv}$ (negative photoconductivity), and enhanced otherwise. By contrast, we systematically observe a positive photoconductivity signal at $\omega_\mathrm{c}\tau<1$ changing to negative at $\omega_\mathrm{c}\tau=1$, which suggests that in all studied Corbino samples heating leads to an increase of the transport time instead of reducing it. In other words, our observations in Corbino samples suggest that $d\tau(T)/dT>0$ at $B\lesssim B_{\rm inv}$.\cite{footnote}

Within the same bolometric mechanism, for the Hall bar measurements one expects no change in the sign of photoresistivity at $B=B_{\rm inv}$. Since the classical Drude resistivity $\rho_{xx}$ is proportional to the momentum relaxation rate $1/\tau$ both in regions of classically weak and classically strong magnetic fields, the sign of bolometric photoresistivity should simply reflect the character of the $T$-dependence of $\tau$. To be consistent with the results obtained on Corbino samples, %(suggesting $d\tau(T)/dT>0$ at $B\lesssim B_{\rm inv}$), {\color{red}I think such relation is excessive and could be even misleading. Idea: to replace is with ''in rather small fields'' or maybe with ''$B < B_{\rm inv}^*$'' Dima Kozlov} 
the photoresistivity measured on Hall bar samples should be negative in some range of low magnetic fields including $B_{\rm inv}$, in full accord with our observations.

We now turn our attention to the second inversion point at $|B|=B_{\rm inv}^*$ where both the photoconductivity in Corbino samples and the photoresistivity in Hall bar samples change sign from negative to positive. Following the same line of reasoning, it is natural to interpret this inversion as transition to a more conventional temperature dependence of the scattering time, such that at $|B|>B_{\rm inv}^*$ it has a negative slope, $d\tau(T)/dT<0$. It is worth mentioning that this inversion corresponds to the parametric region of classically strong magnetic fields, $B_{\rm inv}^*\gg B_{\rm inv}$, 
%{\color{red}Does it correspond to the data. Vasya} 
where $\sigma_{xx}\simeq\rho_{xx}/\rho_{xy}^2$. Since the Hall resistivity $\rho_{xy}\simeq e B/n$ is insensitive to temperature, illumination, and scattering in the relevant classical range of $B$, the qualitative behavior of the photoconductivity and photoresistivity at $B\gg B_{\rm inv}$ should be the same, in accordance with our observations. At the same time, the visibility of the low-$B$ features in photoconductivity (Corbino samples) is strongly enhanced due to the large scaling factor $\rho_{xy}^{-2}\propto B^{-2}$ in comparison with those in the photoresistivity.

To summarize the above discussion, our observations of the sign-alternating photoresponse are consistent with the conventional bolometric effect if one assumes that the momentum relaxation rate decreases (increases) with temperature for $B$ below (above) the value $B=B_{\rm inv}^*$. 

To confirm such an unusual temperature behavior of the momentum scattering rate, we performed additional four-point measurements of the Hall bar resistivity $\rho_{xx}$ at different temperatures in the absence of the terahertz radiation. The results are presented in Fig.~\ref{HallTr}. At first glance, the temperature dependence of $\rho_{xx}$ appears to be rather weak in the relevant classical range of magnetic fields below the onset of Shubnikov-de Haas oscillations. This is consistent with the fact that in the classical region $\rho_{xx}\propto 1/\tau$ is only sensitive to the temperature variations of the transport scattering rate, while the amplitude of Shubnikov-de Haas oscillations includes an explicit and, in our case, much stronger exponential temperature-dependent factor. However, a closer look, see Fig.~\ref{HallTr} (b), demonstrates that the transport rate $1/\tau$ indeed exhibits the temperature dependence that we inferred above from the form of the photoresponse: $1/\tau$ decreases (increases) with temperature for $B$ below (above) $\sim 0.8$~T. This value agrees reasonably well with the values of the high-field inversion point $B_{\rm inv}^*$ in the photoconductivity and photoresistivity thus providing a strong support to the above bolometric interpretation of the sign-alternating photoresponse in our samples.\cite{footnote} At the same time, a surprising result that under some conditions the CR-related \textit{resonant} bolometric response is not detected remains unclear and requires further study.

\section{Summary}
\label{secSUMMARY}

To summarize, our experiments demonstrate that the terahertz analog of the microwave-induced resistance oscillations can be excited in HgTe-based quantum well structures with inverted band order, in spite of their moderate mobility. Furthermore, we found that HgTe structures manifest a distinct sign-alternating photoresponse at low magnetic fields. It features double and single sign inversions in Corbino and Hall bar geometries, respectively. We show that the sign-alternating photoconductivity can be consistently described as the result of electron heating by radiation (bolometric effect). Within this mechanism, the first sign inversion in Corbino samples at $B=B_{\rm inv}$ takes place at $\omega_\mathrm{c}\tau=1$ providing an optoelectronic method to measure the carrier mobility. The second sign inversion at $B=B_{\rm inv}^*$ (common in both geometries) is caused by different temperature behavior of the transport scattering rate at magnetic fields below and above $B=B_{\rm inv}^*$. While this behavior is found to be consistent with the direct transport measurements at different sample temperatures, the effect is much more pronounced in the photoresponse, where it qualitatively modifies the magnetic field dependence. We thus demonstrate that photoconductivity can be a very sensitive probe of the temperature variations of the transport characteristics, even those that are hardly visible using the standard transport measurements. 

\section*{Acknowledgements}
%
%We thank xxxx for discussions. 
The support from the Deutsche Forschungsgemeinschaft (projects  GA501/14-1 and DM1/4-1), and from the Elite Network of Bavaria (K-NW-2013-247) is gratefully acknowledged. Novosibirsk group was supported by the Russian Science Foundation (Grant No. RSF-16-12-10041).


\begin{thebibliography}{99}
\bibitem{footnote}
We mention that bolometric photoconductivity usually measures the sensitivity of the dc transport to the electron temperature $T_\mathrm{e}$, and implies nonequilibrium conditions when $T_\mathrm{e}$ is larger than the lattice (measurement) temperature $T$. By contrast, the $T$-dependent magnetotransport yields information on $\tau(T)$ in conditions of $T_\mathrm{e}=T$. Therefore, the comparison of the results obtained using both methods is in general not straightforward and should be performed with care.
\end{thebibliography}
\end{document}